\documentclass[]{aa}

\usepackage{amsmath}
\usepackage{mathrsfs}
\usepackage{graphicx}
\usepackage{natbib}
\bibpunct{(}{)}{;}{a}{}{,}
\usepackage{breakurl}

\newcommand{\ergs}{\ensuremath{ \mathrm{erg\,s}^{-1} }}

\def\xte{RXTE\xspace}

\def\swift{\textsl{Swift}\xspace}

\newcommand{\inte}{INTEGRAL\xspace}

\def\fu{4U~0115$+$63\xspace}

\def\vo{V0332+53\xspace}
\def\b12{E_{\rm 12}}

\begin{document}

\title{Two giant outbursts of \vo observed with \inte}

\author{
	Carlo Ferrigno\inst{1}
        \and
        Lorenzo Ducci\inst{1,2}
        \and
        Enrico Bozzo\inst{1}
        \and
         Peter Kretschmar\inst{3}
	\and
        Matthias K\"uhnel\inst{4}
	\and
         Christian Malacaria\inst{2}
         \and
         Katja Pottschmidt\inst{5,6}
         \and
         Andrea Santangelo\inst{2}
         \and
         Volodymyr Savchenko\inst{7}
         \and
         J\"orn Wilms\inst{4}
        }

\authorrunning{C. Ferrigno et al.}
\titlerunning{Giant outbursts of \vo}
   \offprints{C. Ferrigno}

\institute{ISDC, University of Geneva, chemin d'\'Ecogia, 16 1290 Versoix Switzerland\\
	\email{carlo.ferrigno@unige.ch}
          \and
          Institut f\"ur Astronomie und Astrophysik, Kepler Center for Astro and Particle Physics, Eberhard Karls Universit\"at, 
              Sand 1, 72076 T\"ubingen, Germany
          \and
          European Space Astronomy Center (ESA/ESAC), Science Operations Department, Villanueva de la Ca\~nada, 28691 Madrid, Spain
           \and
          D Karl Remeis-Sternwarte and Erlangen Center for Astroparticle Physics, Sternwartstr. 7, 96049 Bamberg, Germany
          \and
          CRESST, Department of Physics, and Center for Space Science and
          Technology, UMBC, Baltimore, MD 21250, USA
          \and
          NASA Goddard Space Flight Center, Greenbelt, MD 20771, USA
          \and
          Fran\c{c}ois Arago Center APC - Astroparticule et Cosmologie Universit\'e Paris Diderot, CNRS/IN2P3, CEA/Irfu, Observatoire De Paris, Sorbonne Paris Cit\'e Paris, France
        }

\date{Received ---; accepted ---}

\abstract
	{In July 2015, the high-mass X-ray binary \vo underwent a giant outburst, a decade after the previous one.
         \vo hosts a strongly magnetized neutron star.  During the 2004--2005 outburst, an anti-correlation between the centroid energy
         of its fundamental cyclotron resonance scattering features (CRSFs) and the X-ray luminosity was observed.} 
	{The long ($\approx 100$\,d) and bright ($L_{\rm x} \approx 10^{38}$\,erg\,s$^{-1}$) 2015 outburst 
         offered the opportunity to study during another outburst
         the unique properties of the fundamental CRSF and its dependence on
         the X-ray luminosity.} 
         {The source was observed by the \inte satellite for $\sim 330$\,ks.
          We exploit the spectral resolution at high energies of the SPectrometer on \inte (SPI)
          and the Joint European X-ray Monitors to characterize its spectral properties,
          focusing in particular on the CRSF-luminosity dependence. We complement the data of the
          2015 outburst with those collected by SPI in 2004--2005 and left unpublished so far.} 
         {We find a highly significant anti-correlation of the centroid energy of the fundamental CRSF and the 
         $3-100$\,keV luminosity of $E_1 \propto -0.095(8)L_{37}$\,keV.
         This trend is observed for both outbursts. We confirm the correlation between the width 
         of the fundamental CRSF and the X-ray luminosity previously found in the 
         JEM-X and IBIS dataset of the 2004--2005 outburst.
         By exploiting the RXTE/ASM and Swift/BAT monitoring data we also report on the detection of 
         a $\sim 34$\,d modulation superimposed on the mean profiles and roughly consistent with the orbital period of the pulsar.
         We discuss possible interpretations of
         such variability.}  
	{} 


\maketitle

\section{Introduction}
\label{sect intro}

\vo is a transient X-ray pulsar orbiting around a Be star.
It spends most of the time in a low luminosity state ($L_{\rm x} \la 10^{36} \ergs$),
sporadically interrupted either by normal type I outbursts, 
($L_{\rm x} \sim10^{36-37}\,\ergs$) associated with the passage of the neutron star (NS) at the periastron, 
or by giant type II outbursts, which last several orbital periods.
During these episodes, \vo becomes one of the most luminous X-ray sources of the Galaxy, 
achieving X-ray luminosities up to $L_{\rm x}\approx 10^{38}$\,erg\,s$^{-1}$.
\vo was first detected during a long ($\sim 100$ days) giant outburst caught by Vela~5B in 1973, 
when it reached a peak intensity of 1.4 Crab in $3-12$\,keV ($\sim 2.9 \times 10^{-10}$~erg~cm$^{-2}$~s$^{-1}$; \citealt{terrell1984}). 
A spin period of 4.37\,s and an orbital period of 34\,days (eccentricity $e = 0.37$) 
were discovered \citep{stella1985,zhang2005} during three small outbursts
observed by \emph{EXOSAT} and \emph{Tenma} \citep{Tanaka1983}. 
Owing to the precise source localization obtained with EXOSAT, 
the companion was identified as an O8-9Ve star, BQ Cam \citep{honeycutt1985,Negueruela1999}, and 
its distance estimated to be 2.2--5.8\,kpc \citep{corbet1986}. 
This value was later increased to 7\,kpc
\citep{Negueruela1999}.
\vo was detected again in X-rays by \emph{Ginga} in 1989, leading to the discovery
of an absorption line feature at $\sim$28.5\,keV
and quasi periodic oscillations  at $\sim$0.051\,Hz \citep{Makishima1990, Takeshima1994}.
A particularly bright giant outburst occurred in 2004 November, and was followed 
by \inte and \xte through dedicated target of opportunity observations (TOOs). 
The X-ray spectrum could be fitted using a power law with high-energy cut-off,
a model which is typically adopted to describe the X-ray emission of accreting pulsars 
in Be X-ray binary systems.
Interestingly, the X-ray spectrum shows three absorption-like features 
at energies $\sim28$\,keV, $\sim50$\,keV and $\sim70$\,keV \citep{coburn2005,Pottschmidt2005,kreykenbohm2005}.
These are produced by the resonant scattering of photons on electrons in the accretion column of X-ray pulsars
and are called cyclotron resonance scattering features 
(CRSFs; see, e.g., \citealt{Isenberg1998}; \citealt{Schoenherr2007} and references therein).
If detected, CRSFs provide a direct measurement of the magnetic field of a neutron star
through the relation $E_{\rm cyc} \approx 11.6B_{12}\times(1+z)^{-1}$\,keV,
where $E_{\rm cyc}$ is the centroid energy of the fundamental CRSF, 
$B_{12}$ is the magnetic field strength in units of $10^{12}$\,G, and $z$ is the gravitational redshift of the line-forming region.
The higher harmonics have a centroid energy approximately $n$-times that of the fundamental line.
The equation above can be used to derive a neutron star magnetic field strength of $\sim2.7\times10^{12}$\,G in \vo.
An anti-correlation between the centroid energy of the fundamental CRSF
and the X-ray luminosity during the outbursts has been reported by several authors 
using different data sets \citep{Mihara1998,Mowlavi2006,Tsygankov2006,Tsygankov2010}.
So far, this is the only X-ray pulsar for which an anti-correlation 
is firmly established, as the presence of this phenomenon in
\fu is still debated \citep[see][and references therein]{Mueller2013}. For several other systems,
a positive correlation between X-ray luminosity and CRSF centroid energy has been reported
(Her~X-1, \citealt{Staubert2007}; GX~304$-$1, \citealt{Klochkov2012}; Vela~X-1, \citealt{Furst2013}; 
A\,0535$+$26, \citealt{Sartore2015}).
The anti-correlation observed in \vo has been interpreted as due to an increase of the height
of the accretion column, which induces either the upward migration of 
the line forming region in a region where the magnetic field weakens
\citep[][and references therein]{becker2012},
or the progressive illumination of a larger portion of the neutron star surface,
where cyclotron scattering is assumed to take place in regions progressively further from the magnetic poles \citep{Poutanen2013}.

In June 2015, \vo underwent a new giant outburst \citep{Nakajima2015,Doroshenko2015b},
anticipated by a brightening 
in the optical band, probably associated with the donor star disc \citep{Camero-Arranz2015}. 
During this outburst, INTEGRAL carried out four observations between July 17 and October 9, 
covering both the rise and the decay of the outburst.
In this work, we present the spectral results on the cyclotron line luminosity dependence, 
obtained using the SPectrometer on Integral (SPI) collected during both the source outbursts
in 2004--2005 and 2015.
We complemented the analysis using JEM-X and IBIS data collected during the same periods,
when possible (see the next section).

The observation and data analysis are described in Sect.~\ref{sect. obs}.
In Sect.~\ref{sect. spectral analysis}, we present and discuss the results from the spectral analysis.
In Sect.~\ref{sect. lightcurve}, we report the identification 
of a $\sim 34$\,d modulation superimposed on the profiles of the 2004--2005 and 2015 outbursts.
Finally, in Sect.~\ref{sec:summary}, we summarize our findings.

\section{Observations and data analysis}
\label{sect. obs}
 
The European Space Agency's International Gamma-Ray Astronomy Laboratory (INTEGRAL), 
launched in October 2002, carries three co-aligned coded mask telescopes sensitive in the X- and gamma-ray bands.
We analyze the data from the 
SPectrometer on Integral (SPI; \citealt{Vedrenne03}; \citealt{Roques03}),
the Joint European X-ray Monitor \citep[JEM-X;][]{jemx},
and the Imager on Board the INTEGRAL Satellite \citep[IBIS;][]{ibis}.
The former operates in the 20~keV$-$8~MeV energy range, with a 2$-$8~keV energy
resolution, the second is 
made of two independent units, JEM-X1 and JEM-X2, sensitive from 3 keV to 34 keV, with an energy resolution of 10--15\%,
and the latter operates in the energy range from 15 keV to 600 keV with energy resolution of 6--7\%.

For SPI, we performed the analysis using the SPI Data Analysis
Interface ({\tt spidai}) software provided by the SPI team 
at the IRAP Toulouse\footnote{The software is available at \url{http://sigma-2.cesr.fr/integral/spidai}} 
and the SPI iterative Removal Of Sources ({\tt spiros}) task within the Off-line Science Analysis (OSA) 10.2 software provided
by the ISDC Data Center for Astrophysics (\citealt{Skinner03}; \citealt{Courvoisier03}).
The results obtained with the two softwares have been compared and found to be in good agreement. However, 
owing to a better response characterization below $\sim25$~keV, {\tt spidai}
is better suited for the specific case of \vo, where a broad CRSF at
$\sim27-30$~keV is present. Therefore, only spectra obtained with 
{\tt spidai} are considered in the following analysis. 
Spectra are obtained from a sky model fitting procedure and 
a background based on empty field observations 
(see \citealt{Jourdain09} for a description of the method).
\vo was the only bright X-ray source in the field of view, hence the 
contribution of any other source can be safely neglected in the sky model.
Nonetheless, pointings with unusually high background activity were excluded.

We used OSA 10.2 to obtain the JEM-X spectra in the standard 8 energy bins. By inspecting the time-dependent gain 
parameter file, provided by the instrument team, we discarded the
first science windows of the revolutions in the 2015 data, because of the unstable gain 
evolution after the passage of the spacecraft in the Earth radiation belts. This was not necessary at early times
of the mission, when the calibration sources were strong enough to guarantee a 
reliable energy reconstruction throughout the full satellite revolution.
We limited the analysis to the 3--20\,keV energy range, for which the JEM-X team ensures 
a reliable calibration\footnote{All the information regarding the software and the standard 
calibration can be found on the ISDC web-site at \url{http://isdc.unige.ch/}}. 
In 2004--2005, only the unit JEM-X1 was active, while in 2015, both units were in data-taking mode. 

We reanalyzed the data of the imager IBIS, which appeared already in literature 
\citep{Mowlavi2006,Tsygankov2006} to apply the same model and definition of flux as in the rest of the present paper (see below). 
For the 2015 data, the current version of the software produces IBIS spectra with significant differences from the SPI ones, 
preventing their utilization for the present work.
This is due to the aging of the instrument, which is not optimally accounted for in the software and calibration files, yet.

We analyzed the public data of the TOO observations of the last two outbursts
of \vo, corresponding to the INTEGRAL revolutions 272, 273, 274, 278, 284--288,
1565, 1570, 1596 (no data were available for SPI during revolutions 278 and 1586 because the instrument
was undergoing an annealing cycle, while in revolution 272 the staring pointing mode prevents 
spectral extraction with SPI). The observations were performed with hexagonal dithering 
pattern of the individual $\sim$3-ks-long pointings, therefore all of them fell within the INTEGRAL instrument optimal field of view. 
The log of observations is reported in Table~\ref{tab:observations}. During the spectral fitting, 
for the JEM-X and IBIS spectra, we considered
1\% systematic errors;
we ignored SPI and IBIS data below $\sim 22$\,keV because of uncertainties in the energy response \citep{Jourdain09}; no systematic
uncertainties are added to the SPI spectra.
We used XSPEC version 12.9g \citep{xspec} for the analysis. Uncertainties are reported at
90\% confidence level, unless stated otherwise.

\begin{table*}
\caption{Log of all observations used in this paper.}
\begin{center}
 \begin{tabular}{ l c c c c c}
\hline
\hline
Rev   & Start time [UTC]      & Stop time [UTC] & \multicolumn{3}{c}{Exposures [ks]}\\
      &                       &                       &  JEM-X &  SPI   & IBIS \\
 \hline
272\tablefootmark{a}  &  2005-01-06 05:43:29  &  2005-01-06 18:13:58  &  15.2  &        &  20.1   \\
273  &  2005-01-08 22:18:05  &  2005-01-09 15:45:43  &  35.4  & 44.7   &  38.0  \\
274  &  2005-01-10 03:20:24  &  2005-01-10 09:22:09  &  13.9  & 18.7   &  14.4  \\
278\tablefootmark{b}  &  2005-01-23 15:19:46  &  2005-01-24 15:58:07  &  62.4  &        &  35.3  \\
284  &  2005-02-09 01:14:58  &  2005-02-10 20:43:54  &  127.6  & 124.2 & 100.6  \\
285  &  2005-02-12 01:02:19  &  2005-02-12 05:32:38  &  12.5  & 13.8   &  10.2  \\
286  &  2005-02-15 10:01:35  &  2005-02-15 15:09:16  &  15.0  & 20.6   &  10.2  \\
287  &  2005-02-18 00:36:32  &  2005-02-18 05:31:14  &  13.7  & 15.0   &  11.0  \\
288  &  2005-02-21 00:25:08  &  2005-02-21 05:15:44  &  13.5  & 17.7   &  11.0  \\
1565  &  2015-07-17 05:10:49  &  2015-07-18 03:11:44  &  64.9  & 67.9   &        \\
1570  &  2015-07-30 14:19:15  &  2015-08-01 12:49:39  &  133.8  & 141.2 &        \\
1596  &  2015-10-07 14:13:47  &  2015-10-09 16:53:07  &  145.1  & 120.8 &        \\
\hline
\end{tabular}
\tablefoot{
\tablefoottext{a}{Due to visibility constraints, this observation was performed in ``staring'' mode, for which it is not possible to extract SPI products.}
\tablefoottext{b}{SPI was in its periodic annealing phase and was not taking data.}
}
\label{tab:observations}
\end{center}
\end{table*}

\begin{figure}[]
\begin{center}
\resizebox{\hsize}{!}{\includegraphics[angle=0]{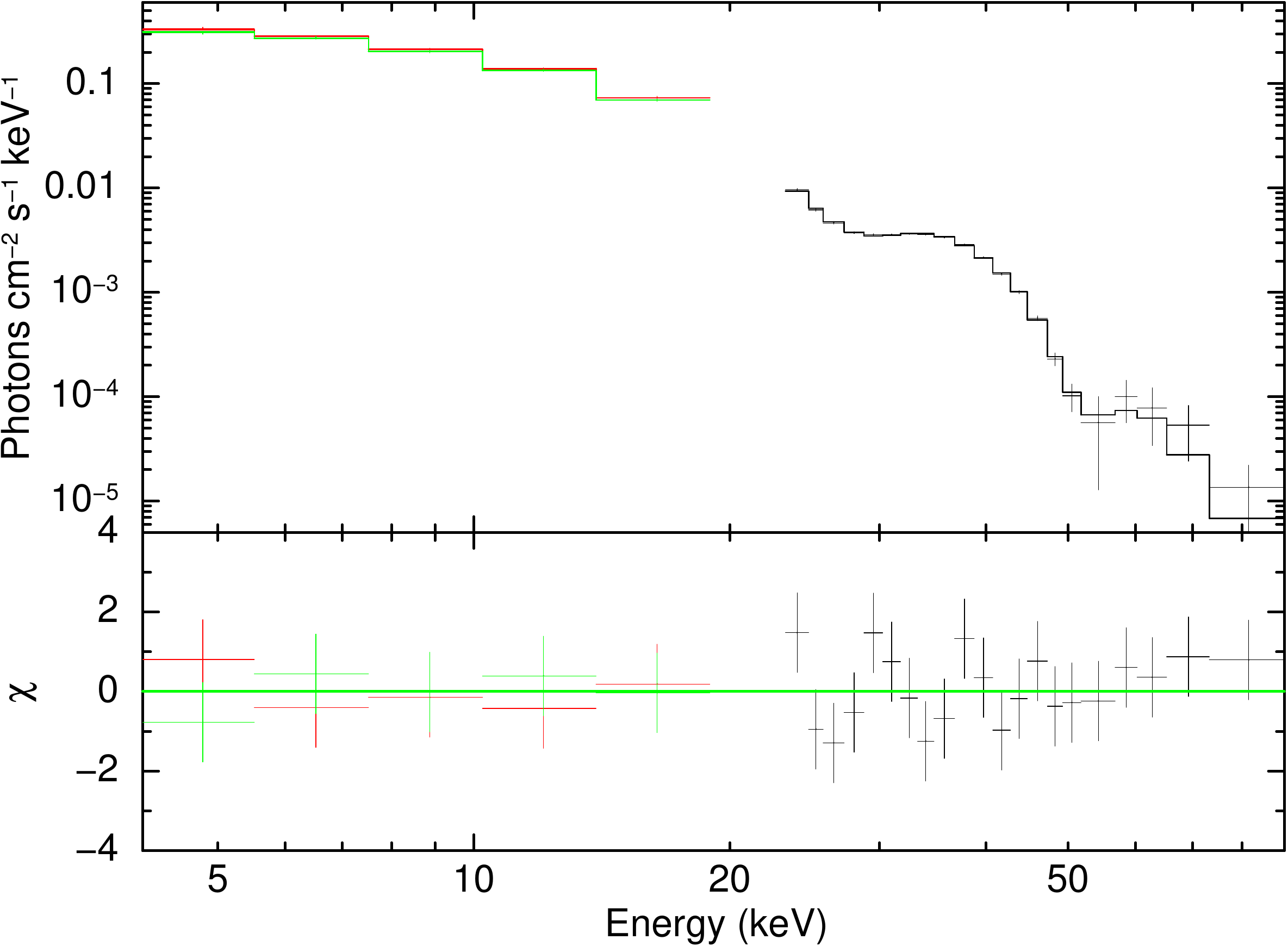}}
\caption{INTEGRAL phase-averaged unfolded spectrum of the \vo observation in satellite revolution 1570. Black, red, and green colors represent SPI, JEM-X1, and JEM-X2 data.}
\label{fig:phase_averaged}
\end{center}
\end{figure}

\section{Spectral analysis}
\label{sect. spectral analysis}

We have modelled the JEM-X+SPI and JEM-X+IBIS spectra using a power law modified 
by an high-energy cutoff and Gaussian shaped absorption profiles for the two observable CRSFs. The spectral model is:
\begin{equation}
N(E)=G_1(E) G_2(E)\left\{
\begin{array}{ll}
E^{-\Gamma}  & \mbox{for } E\le E_C\\
E^{-\Gamma} \exp\left(-\frac{E-E_C}{E_F}\right) & \mbox{for } E>E_C,
\end{array}
\right.
\label{eq:continuum}
\end{equation}
where the functions $G_i(E)$ ($i=1,2$) are the components (\texttt{gabs} model in XSPEC):
\begin{equation}
G_i(E) = \exp \left\{ -\frac{\tau_i}{\sqrt{2\pi}\sigma_i} \exp \left[ -\frac{1}{2}\left(\frac{E-E_{i}}{\sigma_i}\right)^2 \right] \right\}.
\end{equation}
JEM-X data, available only above 3\,keV do not require to include a photoelectric absorption component in the model. 
 
In revolutions 286 and 287 (decline of the outburst) 
the CRSF harmonic was not clearly detected in the SPI spectra
due to the source faintness and the corresponding low statistics of the data.
Therefore, we fixed the centroid energy to 54\,keV in these spectra.
Since the source flux declines exponentially with energy, 
we had to fix the width of the harmonic to $5$\,keV in all the JEM-X plus SPI spectral fits,
to obtain meaningful results.
Due to a degeneracy between the continuum shape and the presence of an absorption features at $\sim$50\,keV in low statistic
spectra, we fitted the JEM-X+IBIS spectra of revolutions 286, 287, 288 with fixed cutoff energy ($14$\,keV) and
second harmonic's width (5\,keV).
Inter-calibration constants were included to account for residual uncertainties in the absolute flux determination,
different exposure times and source variability. We always found values in the expected intervals around unity.

\begin{figure}[]
\begin{center}
\resizebox{\hsize}{!}{\includegraphics[angle=0]{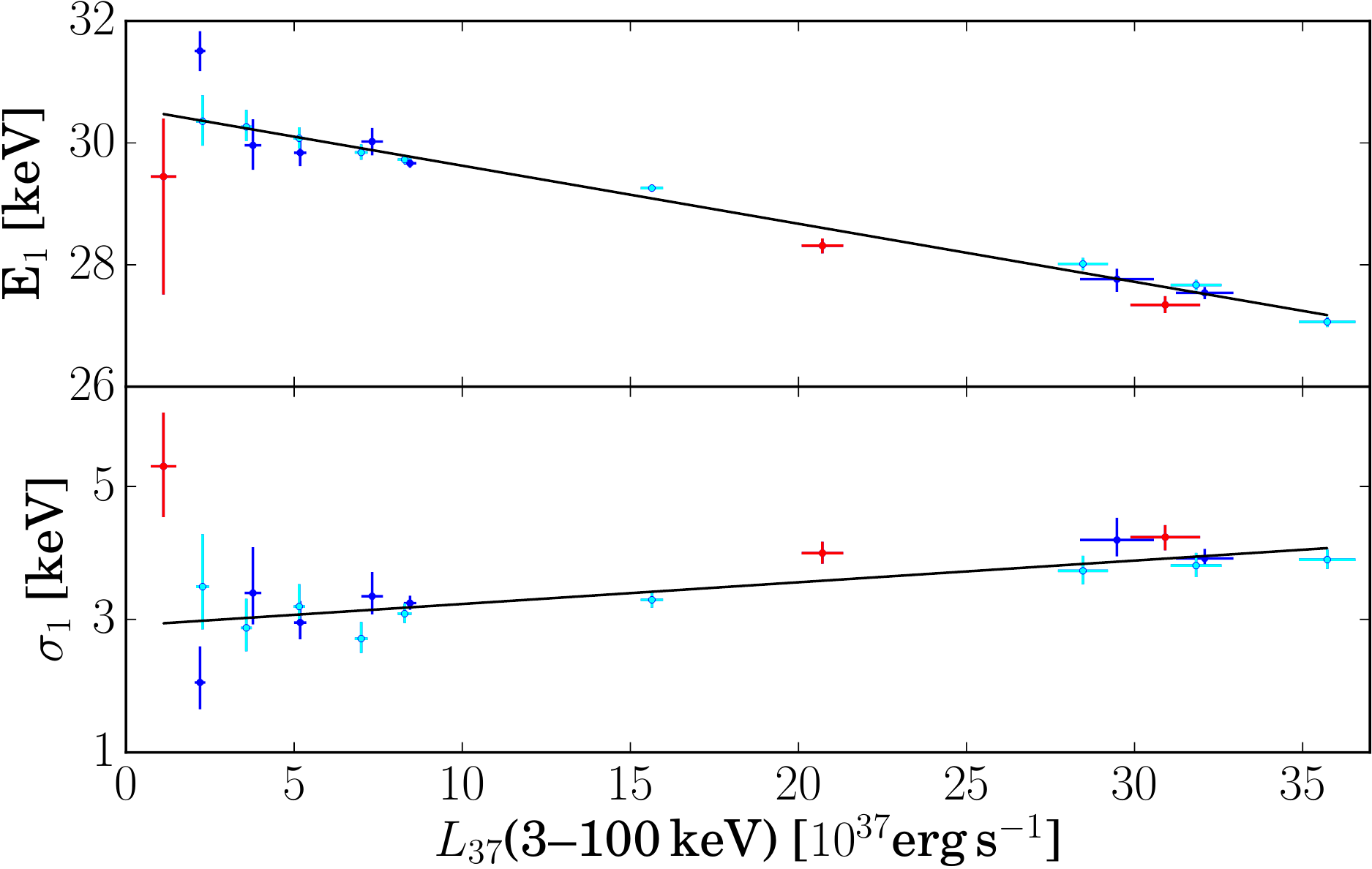}}
\caption{Centroid energy $E_1$ (top panel) and width $\sigma_1$ (bottom panel) of the fundamental CRSF as function of 
the 3-100 keV X-ray luminosity. Red points correspond to the 2015 outburst, 
blue points to the 2004--2005 outburst based on SPI data, and cyan points to the 2004--2005 outburst based on IBIS data.
Black lines show the linear fit. The error bars represent the 68\% confidence intervals.}
\label{fig:lum_resolved}
\end{center}
\end{figure}

The best fit parameters 
are reported in Table \ref{spec results},
while an example of the spectrum during revolution 1570 is shown in Fig.~\ref{fig:phase_averaged}.
The 3--100\,keV luminosity has been calculated using the flux derived
from the best fit spectral continuum model (unaffected by the absorption-like effect of the CRSFs),
assuming a distance of 7\,kpc \citep{Negueruela1999} and isotropic emission.
We note that the slope of the power law and the cutoff energy of revolutions 278, 1565, and 1570 are significantly different from 
the values obtained from the other data sets. In general, there is a strong mathematical degeneracy
between these two parameters for our datasets, but we have verified using confidence contour representations that in all cases,
the statistical improvement is significant. This is suggestive of a subtle difference in the
spectra, which would be anyhow difficult to interpret using phenomenological models. 
On the other hand, the limited statistics of these datasets prevents us from using
more complex, physically motivated models, which have more free parameters.
The parameters describing the continuum are consistent with previous results
(\citealt{Mowlavi2006}; \citealt{Tsygankov2006};  \citealt{Tsygankov2010}).
The centroid energy $E_1$ of the fundamental CRSF decreases by $\sim 14$\%
when the luminosity increases by a factor of  $\sim$30,
while the width $\sigma_1$ of the fundamental CRSF correspondingly increases from $\sim 3$\,keV
up to $\sim 4$\,keV.
We show these trends in Fig.~\ref{fig:lum_resolved}.
We fitted $E_1$ as function of $L_{\rm x}$
and obtained $E_1 = (30.5\pm0.2) - (0.095\pm0.008) L_{\rm 37}$ at 68\% confidence level
for a reduced $\chi^2=2.5$ and 17 degrees of freedom
($L_{37}=L_{\rm x}/10^{37}$\,erg\,s$^{-1}$). 
Fitting the trend $\sigma_1$ vs. $L_{\rm x}$ results in $\sigma_1 = (3.15\pm0.17) +(0.025\pm0.007) L_{\rm 37}$ 
at 68\% confidence level for a reduced $\chi^2=1.9$ and 17 degrees of freedom. Uncertainties are obtained using a 
bootstrap technique with 10\,000 realizations.

\begin{table*}
\caption{Best-fitting parameters for the observations performed during the 2004-2005 and 2015 outbursts.}
\label{spec results}
\centering
\resizebox{\columnwidth+\columnwidth}{!}{%
\begin{tabular}{l r@{}l r@{}l r@{}l r@{}l r@{}l r@{}l r@{}l r@{}l r@{}l r@{}l c}
\noalign{\smallskip}
   \hline
\noalign{\smallskip}
Rev.  & \multicolumn{2}{c}{$\Gamma$}& \multicolumn{2}{c}{$E_{\rm c}$}& \multicolumn{2}{c}{$E_{\rm f}$} & \multicolumn{2}{c}{$E_{\rm{cycl,1}}$} &  \multicolumn{2}{c}{$\sigma_{1}$} & \multicolumn{2}{c}{$\tau_{1}$} & \multicolumn{2}{c}{$E_{\rm{cycl,2}}$} & \multicolumn{2}{c}{$\sigma_{2}$\tablefootmark{a}} & \multicolumn{2}{c}{$\tau_{2}$} & \multicolumn{2}{c}{$L_{\rm x}$ (3--100\,keV)} & $\chi^2_{\rm red}$ (d.o.f.)\\
\noalign{\smallskip}  
      & 	    &               & \multicolumn{2}{c}{keV}      & \multicolumn{2}{c}{keV}        & \multicolumn{2}{c}{keV}            &  \multicolumn{2}{c}{keV}         &         &                      & \multicolumn{2}{c}{keV}           & \multicolumn{2}{c}{keV}          &       &                       &  \multicolumn{2}{c}{$10^{37}$\,erg\,s$^{-1}$} &                          \\
   \hline
\noalign{\smallskip}
  272\,I\tablefootmark{b}& $0.7$&${+0.1\atop-0.1}$   &   $11.2$&${+1.1\atop-1.1}$   & $7.9$&${+0.3\atop-0.3}$          & $27.06$&${+0.13\atop-0.14}$        & $3.9$&${+0.2\atop-0.2}$          & $12.4$&${+1.2\atop-1.0}$       & $50.3$&${+0.5\atop-0.4}$          & $3.1$&${+0.5\atop-0.5}$          & $10.8$&${+1.3\atop -1.1}$    & $35.7$&$\pm0.8$  & 1.073 (36)\\
\noalign{\smallskip}
  273\,S\tablefootmark{b}& $0.8$&${+0.2\atop-0.3}$   &   $12.8$&${+1.3\atop-1.6}$   & $8.9$&${+0.6\atop-0.8}$          & $27.54$&${+0.16\atop-0.17}$        & $3.9$&${+0.2\atop-0.2}$          & $10.7$&${+1.2\atop-1.0}$       & $53.5$&${+1.2\atop-1.1}$          &      &                           & $18$&${+6\atop-4}$           & $32.1$&$\pm0.9$  & 0.742 (17)\\
\noalign{\smallskip}
  273\,I & $0.7$&${+0.1\atop-0.1}$   &   $11.3$&${+1.1\atop-1.1}$   & $8.4$&${+0.3\atop-0.3}$          & $27.67$&${+0.14\atop-0.15}$        & $3.8$&${+0.3\atop-0.3}$          & $11.4$&${+1.3\atop-1.1}$       & $51.6$&${+0.4\atop-0.4}$          & $3.7$&${+0.5\atop-0.5}$          & $13.7$&${+1.5 \atop -1.3}$   & $31.8$&$\pm0.8$  & 0.497 (18)\\ 
\noalign{\smallskip}  
  274\,S & $0.9$&${+0.2\atop-0.3}$   &   $12.4$&${+1.5\atop-2.0}$   & $9.2$&${+0.8\atop-0.8}$          & $27.8$&${+0.3\atop-0.3}$           & $4.2$&${+0.5\atop-0.4}$          & $11.8$&${+3.0\atop-1.9}$       & $52.1$&${+1.7\atop-1.6}$          &      &                          & $19$&${+9\atop-5}$            & $29.5$&$\pm1.1$  & 1.449 (17)\\
\noalign{\smallskip}
  274\,I &$0.59$&${+0.15\atop-0.18}$ &   $10.8$&${+1.3\atop-1.3}$   & $8.3$&${+0.5\atop-0.5}$          & $28.02$&${+0.16\atop-0.17}$        & $3.7$&${+0.4\atop-0.3}$          & $11.4$&${+1.6\atop-1.2}$       & $52.8$&${+0.9\atop-0.8}$          & $4.3$&${+0.9\atop-0.8}$          & $17$&${+4 \atop -3}$        & $28.5$&$\pm0.7$  & 1.228 (18)\\
\noalign{\smallskip}
  278\,I &$-0.40$&${+0.12\atop-0.10}$&   $4.6$&${+0.8\atop-0.9}$    & $6.7$&${+0.3\atop-0.2}$          & $29.26$&${+0.10\atop-0.10}$        & $3.3$&${+0.2\atop-0.2}$          & $11.9$&${+0.8\atop-0.6}$       & $51.7$&${+0.9\atop-0.7}$          & $3.5$&${+0.9\atop-0.8}$         & $15.0$&${+3.2\atop -2.4}$   & $15.6$&$\pm0.3$  & 0.687 (37)\\
\noalign{\smallskip}
  284\,S & $0.66$&${+0.15\atop-0.13}$&   $13.8$&${+1.1\atop-1.1}$   & $9.3$&${+0.5\atop-0.5}$          & $29.67$&${+0.12\atop-0.12}$        &$3.25$&${+0.18\atop-0.16}$        & $11.3$&${+0.8\atop-0.7}$       & $53.3$&${+1.5\atop-1.4}$          &      &                          & $19$&${+9\atop-5}$           & $8.45$&$\pm0.18$ & 1.763 (12)\\
\noalign{\smallskip}
  284\,I & $0.58$&${+0.09\atop-0.10}$&   $13.3$&${+1.0\atop-1.0}$   & $9.3$&${+0.4\atop-0.4}$          & $29.73$&${+0.12\atop-0.11}$        & $3.1$&${+0.3\atop-0.2}$          & $11.9$&${+1.0\atop-0.8}$       & $53.6$&${+0.9\atop-0.8}$          & $5.1$&${+1.0\atop-0.9}$          & $24$&${+6\atop-4}$          & $8.3$&$\pm 0.2$  & 1.176 (18)\\ 
\noalign{\smallskip}
  285\,S & $0.60$&${+0.15\atop-0.19}$& $13.5$&${+1.3\atop-1.5}$     & $9.5$&${+1.1\atop-0.8}$          &  $30.0$&${+0.4\atop-0.4}$           & $3.3$&${+0.6\atop-0.4}$         & $11.8$&${+2.8\atop-1.8}$        &  $53$&${+11\atop-4}$             &      &                          & $19$&${+3474\atop-10}$       & $7.3$&$\pm0.3$   & 1.202 (13)\\
\noalign{\smallskip}
  285\,I & $0.54$&${+0.10\atop-0.12}$& $13.4$&${+1.1\atop-1.1}$     & $8.7$&${+0.5\atop-0.4}$          &  $29.8$&${+0.2\atop-0.2}$           & $2.7$&${+0.4\atop-0.4}$         & $10.7$&${+1.2\atop-0.8}$       & $53.3$&${+2.5\atop-1.6}$          & $3.6$&${+2.1\atop-1.9}$          &  $20$&${+83\atop-7}$        & $7.0$&$\pm0.2$   & 1.067 (18)\\ 
\noalign{\smallskip}
  286\,S & $0.54$&${+0.15\atop-0.17}$& $13.8$&${+1.2\atop-1.2}$     & $8.5$&${+0.7\atop-0.6}$          &  $29.8$&${+0.4\atop-0.4}$           & $3.0$&${+0.5\atop-0.4}$         & $11.5$&${+2.0\atop-1.6}$       &  $54$&\ (fix)                    &              &                  & $36$&${+26 \atop -18}$      & $5.18$&$\pm0.18$ & 1.304 (18)\\
\noalign{\smallskip}
  286\,I & $0.55$&${+0.07\atop-0.07}$&     14&\ (fix)               & $8.6$&${+0.5\atop-0.4}$          &  $30.1$&${+0.3\atop-0.3}$           & $3.2$&${+0.6\atop-0.5}$         & $11.8$&${+1.8\atop-1.3}$       &     56 & $^{+7}_{-3}$    &      &                          &     44 & $^{+941}_{-22}$                       & $5.15$&$\pm0.17$  & 1.180  (20)\\ 
\noalign{\smallskip}
  287\,S & $0.63$&${+0.15\atop-0.15}$& $15.1$&${+2.3\atop-1.4}$     & $9.0$&${+1.3\atop-1.0}$          &  $30.0$&${+0.7\atop-0.7}$           & $3.4$&${+1.1\atop-0.8}$         & $11.6$&${+0.5\atop-2.9}$       & $54$&\ (fix)                     &       &                         & $24$&${+40\atop-21}$         & $3.78$&$\pm0.25$ & 1.461 (18)\\
\noalign{\smallskip}
  287\,I & $0.54$&${+0.07\atop-0.07}$&     14&\ (fix)               &  $8.9$&${+1.1\atop-0.7}$         &  $30.3$&${+0.5\atop-0.4}$           & $2.9$&${+0.7\atop-0.6}$         & $11.9$&${+2.8\atop-1.6}$       &      56 & $^{+9}_{-3}$                        &        &                        &     $>$30&                       & $3.58$&$\pm0.16$   & 1.389 (20)\\ 
\noalign{\smallskip}
  288\,S & $0.68$&${+0.16\atop-0.18}$&   $14.7$&${+2.3\atop-1.6}$   & $9.5$&${+3.3\atop-1.6}$          &  $31.5$&${+0.5\atop-0.5}$           & $2.1$&${+0.9\atop-0.6}$         & $12.1$&${+11.2\atop-3.7}$      & $57$&${+13\atop-7}$              &      &                         & $464$&${+464\atop-439}$    & $2.20$&$\pm0.16$ & 1.198 (17)\\
\noalign{\smallskip}
  288\,I & $0.66$&${+0.08\atop-0.08}$&     14&\ (fix)               & $9.1$&${+1.7\atop-1.1}$          & $30.4$&${+0.7\atop-0.7}$            & $3.5$&${+1.3\atop-1.1}$         & $13.0$&${+6.2\atop-3.5}$       &     51 & $^{+10}_{-3}$                            &       &                       &    15 & $^{+62}_{-11}$                      & $2.28$&$\pm0.19$ & 1.358  (13)\\ 
\noalign{\smallskip}
 1565\,S & $-0.30$&${+0.18\atop-0.18}$&  $5.3$&${+0.7\atop-1.6}$    & $6.7$&${+0.6\atop-0.6}$          &  $28.31$&${+0.19\atop-0.21}$        & $4.0$&${+0.3\atop-0.3}$         & $13.4$&${+2.0\atop-1.7}$       & $52.5$&${+1.4\atop-1.2}$         &       &                          & $16$&${+5.8\atop-3.8}$     & $20.7$&$\pm0.6$  & 1.339 (21)\\
\noalign{\smallskip}
 1570\,S & $-0.20$&${+0.27\atop-0.25}$&  $5.2$&${+1.0\atop-5.2}$    & $6.5$&${+1.0\atop-0.7}$          &  $27.34$&${+0.23\atop-0.22}$        & $4.2$&${+0.3\atop-0.3}$         & $13.0$&${+2.5\atop-2.4}$       & $53.9$&${+1.5\atop-1.2}$         &       &                          & $21$&${+10\atop-6}$         & $30.9$&$\pm1.0$  & 1.166 (20)\\
\noalign{\smallskip}
 1596\,S & $0.81$&${+0.06\atop-0.08}$&  $17$&${+16\atop-2.8}$       & $18$&${+7\atop-5}$               &  $29.4$&${+1.6\atop-3.2}$          &  $5.3$&${+1.3\atop-1.3}$         & $19$&${+20\atop-9}$            & $53$&${+4\atop-4}$              &        &                          & $383$&${+6128\atop-314}$    & $1.1$&$\pm0.4$   & 1.261 (21) \\
\noalign{\smallskip}
   \hline
\end{tabular}
}
\tablefoot{
\tablefoottext{a}{The width $\sigma_{2}$ of the first harmonic of the CRSFs is fixed to $5$\,keV unless indicated otherwise.}
\tablefoottext{b}{Rows with the label ``S'' refer to the simultaneous JEM-X and SPI data; rows with the label ``I'' to the JEM-X and IBIS data.}
}
\end{table*}

The three new measurements of $E_1$ during the 2015 outburst (red points in Fig. \ref{fig:lum_resolved})
nicely follow the trend of the previous outburst, although they lie below the best-fit linear correlation at
a $\sim$2$\sigma$ level.
Combined with the similarity in the spectral continuum, 
this confirms also for this source similar accretion properties
over different outbursts. This is rather typical of Be/X-ray binary systems, see the cases of 
4U 0115+63 \citep[][, and references therein]{Mueller2013}, EXO\,2030+375 \citep{naik2013,wilson2008},
XTE~J1946+274 \citep{Mueller2012}, and GRO~J1008$-$57 \citep{Kuehnel2013}.
The anti-correlation of $E_1$ with $L_{\rm x}$ obtained from the full data set and 
separately from the 2004--2005 and 2015 data subsets confirms the trend previously reported for this source.
\citet{Tsygankov2006} fitted the CRSFs with the cyclotron absorption line model of \citet{Mihara1990,Makishima1990}
({\tt cyclabs} in XSPEC). It is known that the centroid energies obtained with this model
    differ from that obtained with the {\tt gabs} model \citep[see, e.g.,][]{Suchy2011}.
    Therefore, a comparison between the slope of $E_1$ vs. $L_X$ obtained in this work with
    that of \citet{Tsygankov2006} cannot be carried out.
\citet{Cusumano2016} studied the evolution of the fundamental CRSF during the 2015 outburst of \vo
using \swift data. 
Our measurements of $E_1$ are consistent with those of \citet{Cusumano2016} if the latter are shifted in energy by $\sim 1$\,keV.
Such discrepancy is most likely due to calibration systematics between the two instruments.

Moreover, \citet{Cusumano2016} found that the $E_1$ vs. $L_{37}$ slopes in the rise and decay phases of the outburst are 
significantly different, causing an increase of $\sim 1.5$\,keV of the cyclotron energy $E_1$
from the onset to the end of the outburst.
This would correspond to  a drop of about $1.7\times 10^{11}$\,G of the
magnetic field in the photon emitting region. 
In our dataset, we have only three points, therefore our sensitivity to different trends is limited.
By assuming that the point extracted in the satellite revolution 1570 corresponds to the peak of the outburst, 
we have computed slopes in the ascending  and descending branches (revolutions 1565-1570 and 1570-1596, respectively).
Both values are compatible with our average value
within an uncertainty of 0.03 keV/$L_{37}$ (68\% confidence level). 
In \citet{Cusumano2016},  the slopes are $-0.137\pm$0.008 and $-0.074\pm0.005$ keV/$L_{37}$ 
for the ascending and descending branches, respectively. 
Our measurement, dominated by the 2004--2005 data, lies between them, while
our determinations based on SPI data of the 2015 outburst are consistent with both values at a 2$\sigma$ level,
showing that these data are not sufficient for an independent measurement.
The inclusion of IBIS for the 2015 outburst when a refined energy calibration will be available, 
could possibly allow us to make a better comparison.

We confirm the trend of the absorption line width versus luminosity
found by \citet{Mowlavi2006} in the JEM-X and IBIS data collected during the 2004--2005 outburst.
This is expected if the emission region broadens with luminosity due to the growth of the accretion column,
bringing a larger and larger range of magnetic field intensities into the emission region.
However, \citet{Tsygankov2010} do not detect a significant correlation of the absorption line width versus luminosity.
By inspecting their Fig.~5, we notice that
significant trends could be obtained by isolating some subsets of data; this might suggest that the 
smaller INTEGRAL sample could be biased.
However, by using the Gaussian absorption model on the Swift/BAT data covering both outburst, a positive correlation
of line width versus luminosity is 
unambiguously detected (Cusumano, private communication).
We conclude that the use of different spectral models is the most 
likely cause of this apparent contradiction with \citet{Tsygankov2010}.

In the model of \citet{Poutanen2013}, an increase of the line width is expected alongside the variation of the
centroid energy due to the enlarged region of X-ray reflection (panels a and b of their Fig.~4), while \citet{Nishimura2014} argue that the lines are formed 
on the column's side walls in relatively small portions with nearly constant size. He infers that
the height on the NS surface is so high in \vo that the effects of bulk motion become weak in focusing emission
towards the NS surface, leading to a negligible reflection.
This could also explain the general constancy of hard X-ray continuum parameters (Table~\ref{spec results}), which are expected 
to track the temperature and geometrical configuration of the emitting plasma. These should not dramatically vary with luminosity, 
when the emitting region is far away from the radiation shock. 
Unfortunately, the details of why these features
are so widely variable between sources are largely unknown; they rely on long-sought theoretical developments.

\begin{figure*}
\begin{center}
\includegraphics[width=\columnwidth]{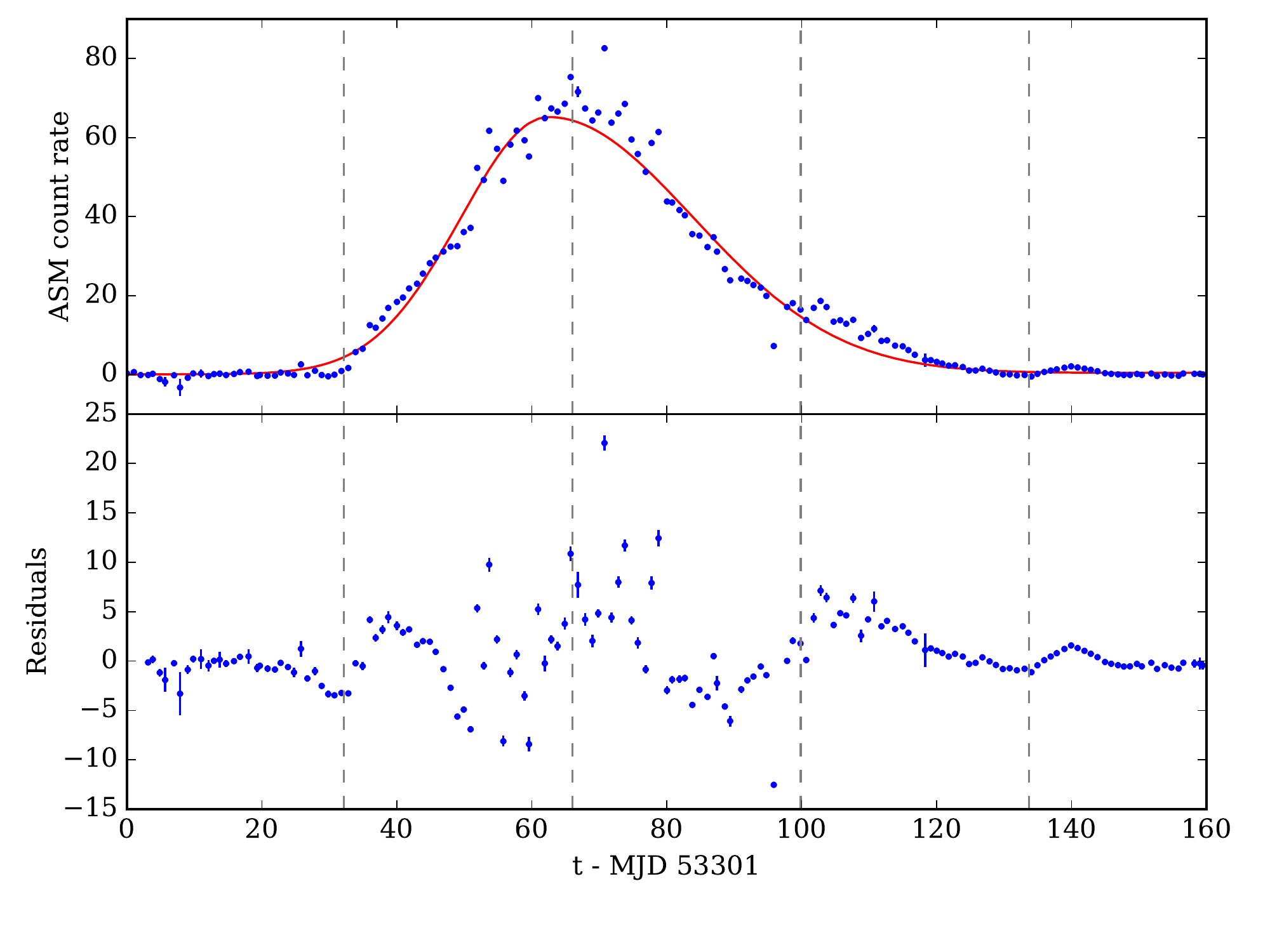}
\includegraphics[width=\columnwidth]{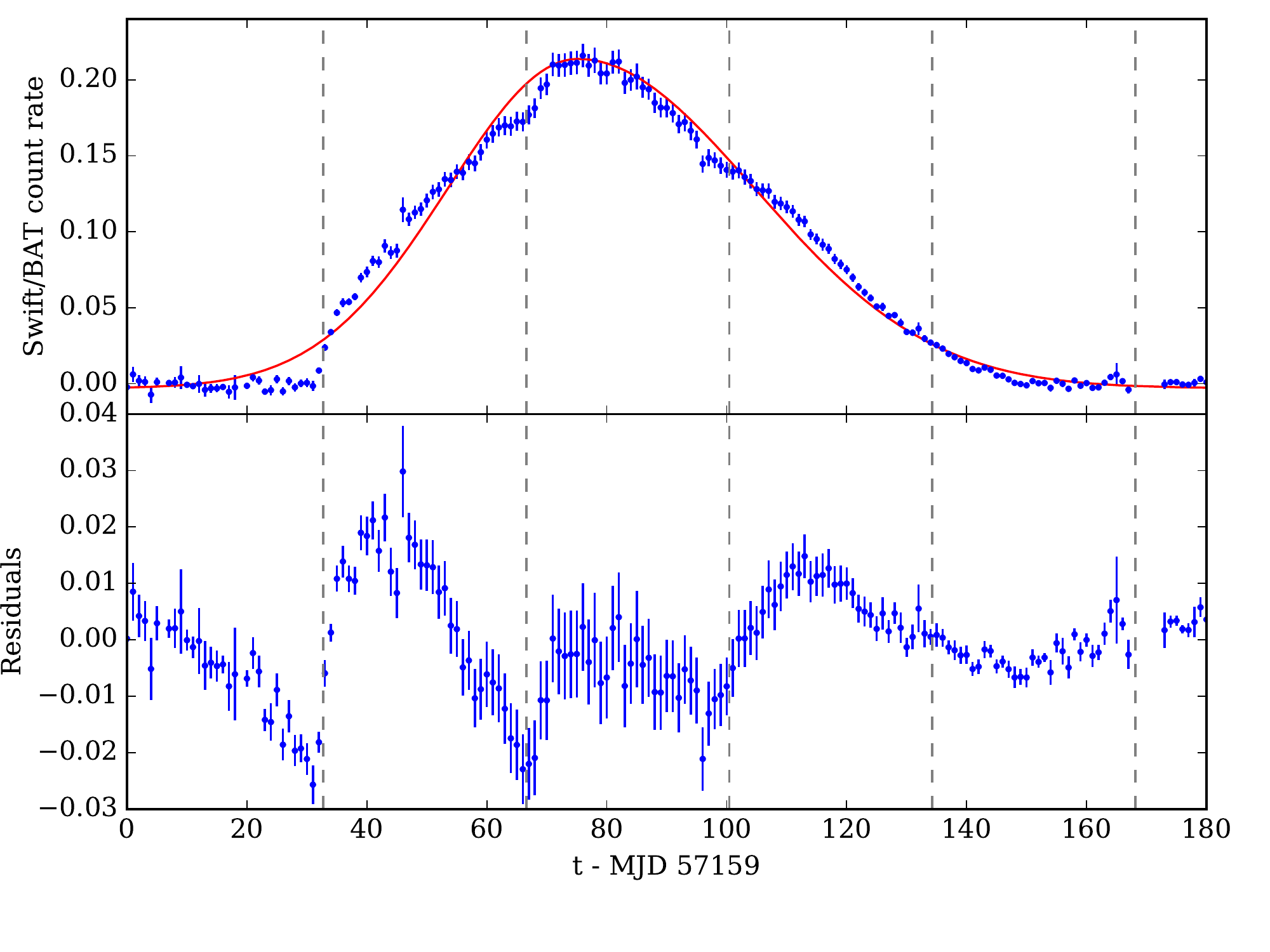}
\caption{\emph{Top panels:} outburst profiles (blue points) of \vo observed by RXTE/ASM in 2004--2005 (\emph{left panel})
and by \emph{Swift}/BAT in 2015 (\emph{right panel}) fitted with the asymmetric Gaussian function described in the text (red lines).
\emph{Bottom panels:} detrended data showing the $\sim34$\,d orbital modulation.}
\label{fig:outb_profiles}
\end{center}
\end{figure*}

\section{Outburst profiles}
\label{sect. lightcurve}

The giant outburst of \vo in 1973 June-September showed 
a modulation at $\sim 34$ days,
which allowed \citet{Whitlock1989} to refine the measurement of the orbital period reported by \citet{stella1985}.
To our knowledge, this is the only report of an orbital modulation 
superimposed on a giant outburst of a Be/X-ray binary (see \citealt{Okazaki2013}).
Also the Be/XRB SAX~J2103.5$+$4545 showed a correlation
of the X-ray luminosity with the orbital period ($\sim 12.7$\,d) during the
bright states of 2010 and 2012 \citep{Camero2014}.
However, the outbursts of SAX~J2103.5$+$4545 were fainter ($L_{\rm x} \la 10^{37}$\,erg\,s$^{-1}$)
compared to \vo and their properties are reminiscent of type I outbursts.
To explore whether such a behavior is recurrent in \vo, we retrieved the daily-averaged lightcurves obtained 
by the All Sky Monitor (ASM, 2$-$10\,keV) on board Rossi X-ray Timing Explorer (RXTE) satellite
and the Burst Alert Telescope on board Swift \citep[BAT, 15$-$50\,keV][]{Krimm2013}\footnote{Provided by the ASM/RXTE (\url{https://heasarc.gsfc.nasa.gov/docs/xte/asm_products.html}) and \swift/BAT teams (\url{http://swift.gsfc.nasa.gov/results/transients/index.html}).}.
The profiles of the 2004--2005 and 2015 giant outbursts of \vo are shown in the upper panels of Fig.~\ref{fig:outb_profiles}.
The 2015 episode lasted $\sim$110 days, slightly longer than the 2004--2005 outburst ($\sim 90$ days) 
and both reached an X-ray luminosity of few $10^{38}$\,erg\,s$^{-1}$.
Both the curves seem to show a modulation of $\sim 34$~days 
superimposed on the rising and decaying phases.
To highlight such variability, we fitted the outbursts with an asymmetric Gaussian function and a constant 
(red lines in top panels of Fig. \ref{fig:outb_profiles}; 
see \citealt{Kuehnel2015}, for details about the asymmetric Gaussian model).
Then, we subtracted the fitting functions from the outburst profiles and we plotted the 
resulting detrended lightcurves (bottom panels of Fig.~\ref{fig:outb_profiles}).
Dashed vertical lines show the times of periastron passages, obtained
from the ephemeris of \citet{Doroshenko2015}.
We found that during the two outbursts, the  sine-like modulation has 
maxima $\approx$10--15\,days after the periastron passages.
An enhancement of the X-ray luminosity that repeats periodically
at intervals of $P_{\rm orb}$ in wind-fed systems with high eccentricity
can be explained with the higher accretion rate at the periastron passage of the pulsar,
where the circumstellar wind is denser and slower (e.g., \citealt{Waters1989}; \citealt{Raguzova1998}).
An accretion disk around the NS of \vo is expected to form during giant outbursts as
indicated by the strong spin-up of the NS \citep{Doroshenko2015}.
The timescale needed for the gravitationally captured material to spiral-in and 
fall onto the NS is of the order of $\ga 10-100$ days, and depends on the 
dimensions of the disk and its viscosity properties \citep{Shakura1973}.
\citet{Waters1989} pointed out that if this timescale is not too large,
the X-ray modulation caused by the accretion rate variability along the
eccentric orbit of the neutron star is not completely flattened
and could produce a modulation, which is variable on the orbital timescale.
Since the accretion disk behaves like a reservoir of matter,
the enhanced amount of matter gravitationally captured at periastron
is not immediately accreted on the surface of the neutron star,
hence a phase shift between the periastron passage and the maximum of the X-ray luminosity
as that observed in Fig.~\ref{fig:outb_profiles} is expected. While this qualitative explanation 
is plausible, more detailed calculations are required for a quantitative estimate, which is outside the scope of the present work.

\section{Summary}
\label{sec:summary}
We have reported results of the spectral properties of \vo during
the 2004--2005 and 2015 giant outbursts, based on SPI, JEM-X, and IBIS observations.
We showed that the spectral parameters describing the continuum are consistent
with previous results.
The correlation of the centroid of the fundamental CRSF and the 3$-$100\,keV
luminosity is $E_1 \propto -0.095\pm 0.008 L_{37}$\,keV.
The value of the slope lies between the slopes determined by \citet{Cusumano2016} for
the ascending and descending phases of the 2015 outburst, while a direct comparison with the results
by \citet{Tsygankov2006} is hampered by the different spectral model adopted by these authors.
Using SPI data of the 2004--2005 outburst and left unpublished so far
together with JEM-X and IBIS data, we confirmed a significant correlation between the 
absorption line width of the fundamental CRSF and the 3--100\,keV luminosity
previously found by \citet{Mowlavi2006} in the JEM-X+ISGRI dataset of the same outburst.
We found a modulation at $\sim34$ days superimposed on the 2004--2005 and 2015 outburst profiles
in RXTE ASM and \swift/BAT lightcurves, with maxima shifted of 10$-$15 days after
the periastron passage.
Such modulation in the lightcurve of a neutron star powered by an accretion disk
can be explained with an enhanced amount of matter gravitationally captured
at periastron and spiraling inwards (on a timescale that depends on the dimension
of the disk and its viscosity properties) until it is accreted on the surface of the neutron star.

\begin{acknowledgements}
CF thanks G. Cusumano for his constructive comments on our manuscript and 
for discussing Swift/BAT results.
LD acknowledges Jean-Pierre Roques, Elisabeth Jourdain, and James Rodi for their
precious help on the use of SPIDAI.
LD and MK acknowledge support by the Bundesministerium f\"ur Wirtschaft und Technologie and
the Deutsches Zentrum f\"ur Luft und Raumfahrt through the grants FKZ 50 OG 1602 and 50 OR 1207, respectively.
This paper is based on data from observations with INTEGRAL, 
an ESA project with instruments and science data centre funded by ESA
member states (especially the PI countries: Denmark, France, Germany,
Italy, Spain, and Switzerland), Czech Republic and Poland,
and with the participation of Russia and the USA.
\end{acknowledgements}

\bibliographystyle{aa}
\bibliography{references}

\begin{thebibliography}{53}
\expandafter\ifx\csname natexlab\endcsname\relax\def\natexlab#1{#1}\fi

\bibitem[{Arnaud(1996)}]{xspec}
Arnaud, K.~A. 1996, Astronomical Data Analysis Software and Systems V, 101, 17

\bibitem[{Becker {et~al.}(2012)Becker, Klochkov, Sch{\"o}nherr, Nishimura,
  Ferrigno, Caballero, Kretschmar, Wolff, Wilms, \& Staubert}]{becker2012}
Becker, P.~A., Klochkov, D., Sch{\"o}nherr, G., {et~al.} 2012, \aap, 544, 123

\bibitem[{{Camero} {et~al.}(2014){Camero}, {Zurita}, {Guti{\'e}rrez-Soto},
  {{\"O}zbey Arabac{\i}}, {Nespoli}, {Kiaeerad}, {Beklen},
  {Garc{\'{\i}}a-Rojas}, \& {Caballero-Garc{\'{\i}}a}}]{Camero2014}
{Camero}, A., {Zurita}, C., {Guti{\'e}rrez-Soto}, J., {et~al.} 2014, \aap, 568,
  A115

\bibitem[{{Camero-Arranz} {et~al.}(2015){Camero-Arranz}, {Caballero-Garcia},
  {Ozbey-Arabaci}, \& {Zurita}}]{Camero-Arranz2015}
{Camero-Arranz}, A., {Caballero-Garcia}, M., {Ozbey-Arabaci}, M., \& {Zurita},
  C. 2015, ATEL, 7682

\bibitem[{{Coburn} {et~al.}(2005){Coburn}, {Kretschmar}, {Kreykenbohm},
  {McBride}, {Rothschild}, \& {Wilms}}]{coburn2005}
{Coburn}, W., {Kretschmar}, P., {Kreykenbohm}, I., {et~al.} 2005, ATEL, 381

\bibitem[{{Corbet} {et~al.}(1986){Corbet}, {Charles}, \& {van der
  Klis}}]{corbet1986}
{Corbet}, R.~H.~D., {Charles}, P.~A., \& {van der Klis}, M. 1986, \aap, 162,
  117

\bibitem[{{Courvoisier} {et~al.}(2003){Courvoisier}, {Walter}, {Beckmann},
  {Dean}, {Dubath}, {Hudec}, {Kretschmar}, {Mereghetti}, {Montmerle},
  {Mowlavi}, {Paltani}, {Preite Martinez}, {Produit}, {Staubert}, {Strong},
  {Swings}, {Westergaard}, {White}, {Winkler}, \& {Zdziarski}}]{Courvoisier03}
{Courvoisier}, T.~J.-L., {Walter}, R., {Beckmann}, V., {et~al.} 2003, \aap,
  411, L53

\bibitem[{{Cusumano} {et~al.}(2016){Cusumano}, {La Parola}, {D'Ai}, {Segreto},
  {Tagliaferri}, {Barthelmy}, \& {Gehrels}}]{Cusumano2016}
{Cusumano}, G., {La Parola}, V., {D'Ai}, A., {et~al.} 2016, \aap in press
  [\eprint[arXiv]{1604.07831}]

\bibitem[{{Doroshenko} {et~al.}(2015){Doroshenko}, {Tsygankov}, {Ferrigno},
  {Bozzo}, {Lutovinov}, \& {Mushtukov}}]{Doroshenko2015b}
{Doroshenko}, V., {Tsygankov}, S., {Ferrigno}, C., {et~al.} 2015, ATEL, 7822

\bibitem[{{Doroshenko} {et~al.}(2016){Doroshenko}, {Tsygankov}, \&
  {Santangelo}}]{Doroshenko2015}
{Doroshenko}, V., {Tsygankov}, S., \& {Santangelo}, A. 2016, \aap, 589, A72

\bibitem[{F{\"u}rst {et~al.}(2013)F{\"u}rst, Pottschmidt, Wilms, Tomsick,
  Bachetti, Boggs, Christensen, Craig, Grefenstette, Hailey, Harrison, Madsen,
  Miller, Stern, Walton, \& Zhang}]{Furst2013}
F{\"u}rst, F., Pottschmidt, K., Wilms, J., {et~al.} 2013, \apj, 780, 133

\bibitem[{{Honeycutt} \& {Schlegel}(1985)}]{honeycutt1985}
{Honeycutt}, R.~K. \& {Schlegel}, E.~M. 1985, \pasp, 97, 300

\bibitem[{{Isenberg} {et~al.}(1998){Isenberg}, {Lamb}, \&
  {Wang}}]{Isenberg1998}
{Isenberg}, M., {Lamb}, D.~Q., \& {Wang}, J.~C.~L. 1998, \apj, 505, 688

\bibitem[{{Jourdain} \& {Roques}(2009)}]{Jourdain09}
{Jourdain}, E. \& {Roques}, J.~P. 2009, \apj, 704, 17

\bibitem[{Klochkov {et~al.}(2012)Klochkov, Doroshenko, Santangelo, Staubert,
  Ferrigno, Kretschmar, Caballero, Wilms, Kreykenbohm, Pottschmidt, Rothschild,
  Wilson-Hodge, \& P{\"u}hlhofer}]{Klochkov2012}
Klochkov, D., Doroshenko, V., Santangelo, A., {et~al.} 2012, \aap, 542, L28

\bibitem[{{Kreykenbohm} {et~al.}(2005){Kreykenbohm}, {Mowlavi}, {Produit},
  {Soldi}, {Walter}, {Dubath}, {Lubi{\'n}ski}, {T{\"u}rler}, {Coburn},
  {Santangelo}, {Rothschild}, \& {Staubert}}]{kreykenbohm2005}
{Kreykenbohm}, I., {Mowlavi}, N., {Produit}, N., {et~al.} 2005, \aap, 433, L45

\bibitem[{{Krimm} {et~al.}(2013){Krimm}, {Holland}, {Corbet}, {Pearlman},
  {Romano}, {Kennea}, {Bloom}, {Barthelmy}, {Baumgartner}, {Cummings},
  {Gehrels}, {Lien}, {Markwardt}, {Palmer}, {Sakamoto}, {Stamatikos}, \&
  {Ukwatta}}]{Krimm2013}
{Krimm}, H.~A., {Holland}, S.~T., {Corbet}, R.~H.~D., {et~al.} 2013, \apjs,
  209, 14

\bibitem[{{K{\"u}hnel} {et~al.}(2015){K{\"u}hnel}, {Kretschmar}, {Nespoli},
  {Okazaki}, {Schoenherr}, {Wilson-Hodge}, {Falkner}, {Brand}, {Anders},
  {Schwarm}, {Kreykenbohm}, {M{\"u}ller}, {Pottschmidt}, {Fuerst}, {Grinberg},
  \& {Wilms}}]{Kuehnel2015}
{K{\"u}hnel}, M., {Kretschmar}, P., {Nespoli}, E., {et~al.} 2015, in
  Proceedings of ''A Synergistic View of the High Energy Sky'' - 10th INTEGRAL
  Workshop (INTEGRAL 2014). 15-19 September 2014. Annapolis, MD, USA., 78

\bibitem[{{K{\"u}hnel} {et~al.}(2013){K{\"u}hnel}, {M{\"u}ller}, {Kreykenbohm},
  {F{\"u}rst}, {Pottschmidt}, {Rothschild}, {Caballero}, {Grinberg},
  {Sch{\"o}nherr}, {Shrader}, {Klochkov}, {Staubert}, {Ferrigno},
  {Torrej{\'o}n}, {Mart{\'{\i}}nez-N{\'u}{\~n}ez}, \& {Wilms}}]{Kuehnel2013}
{K{\"u}hnel}, M., {M{\"u}ller}, S., {Kreykenbohm}, I., {et~al.} 2013, \aap,
  555, A95

\bibitem[{Lund {et~al.}(2003)Lund, Budtz-J{\o}rgensen, Westergaard, Brandt,
  Rasmussen, Hornstrup, Oxborrow, Chenevez, Jensen, Laursen, Andersen,
  Mogensen, Rasmussen, Om{\o}, Pedersen, Polny, Andersson, Andersson,
  K{\"a}m{\"a}r{\"a}inen, Vilhu, Huovelin, Maisala, Morawski, Juchnikowski,
  Costa, Feroci, Rubini, Rapisarda, Morelli, Carassiti, Frontera, Pelliciari,
  Loffredo, Mart~inez N{\'u}{\~n}ez, Reglero, Velasco, Larsson, Svensson,
  Zdziarski, Castro-Tirado, Attina, Goria, Giulianelli, Cordero, Rezazad,
  Schmidt, Carli, Gomez, Jensen, Sarri, Tiemon, Orr, Much, Kretschmar, \&
  Schnopper}]{jemx}
Lund, N., Budtz-J{\o}rgensen, C., Westergaard, N.~J., {et~al.} 2003, \aap, 411,
  L231

\bibitem[{{Makishima} {et~al.}(1990){Makishima}, {Mihara}, {Ishida}, {Ohashi},
  {Sakao}, {Tashiro}, {Tsuru}, {Kii}, {Makino}, {Murakami}, {Nagase}, {Tanaka},
  {Kunieda}, {Tawara}, {Kitamoto}, {Miyamoto}, {Yoshida}, \&
  {Turner}}]{Makishima1990}
{Makishima}, K., {Mihara}, T., {Ishida}, M., {et~al.} 1990, \apjl, 365, L59

\bibitem[{{Mihara} {et~al.}(1998){Mihara}, {Makishima}, \&
  {Nagase}}]{Mihara1998}
{Mihara}, T., {Makishima}, K., \& {Nagase}, F. 1998, Advances in Space
  Research, 22, 987

\bibitem[{{Mihara} {et~al.}(1990){Mihara}, {Makishima}, {Ohashi}, {Sakao}, \&
  {Tashiro}}]{Mihara1990}
{Mihara}, T., {Makishima}, K., {Ohashi}, T., {Sakao}, T., \& {Tashiro}, M.
  1990, \nat, 346, 250

\bibitem[{Mowlavi {et~al.}(2006)Mowlavi, Kreykenbohm, Shaw, Pottschmidt, Wilms,
  Rodriguez, Produit, Soldi, Larsson, \& Dubath}]{Mowlavi2006}
Mowlavi, N., Kreykenbohm, I., Shaw, S.~E., {et~al.} 2006, \aap, 451, 187

\bibitem[{M{\"u}ller {et~al.}(2013)M{\"u}ller, Ferrigno, K{\"u}hnel,
  Sch{\"o}nherr, Becker, Wolff, Hertel, Schwarm, Grinberg, Obst, Caballero,
  Pottschmidt, F{\"u}rst, Kreykenbohm, Rothschild, Hemphill, N{\'u}{\~n}ez,
  Torrej{\'o}n, Klochkov, Staubert, \& Wilms}]{Mueller2013}
M{\"u}ller, S., Ferrigno, C., K{\"u}hnel, M., {et~al.} 2013, \aap, 551, 6

\bibitem[{{M{\"u}ller} {et~al.}(2012){M{\"u}ller}, {K{\"u}hnel}, {Caballero},
  {Pottschmidt}, {F{\"u}rst}, {Kreykenbohm}, {Sagredo}, {Obst}, {Wilms},
  {Ferrigno}, {Rothschild}, \& {Staubert}}]{Mueller2012}
{M{\"u}ller}, S., {K{\"u}hnel}, M., {Caballero}, I., {et~al.} 2012, \aap, 546,
  A125

\bibitem[{{Naik} {et~al.}(2013){Naik}, {Maitra}, {Jaisawal}, \&
  {Paul}}]{naik2013}
{Naik}, S., {Maitra}, C., {Jaisawal}, G.~K., \& {Paul}, B. 2013, \apj, 764, 158

\bibitem[{{Nakajima} {et~al.}(2015){Nakajima}, {Mihara}, {Negoro}, {Kawai},
  {Ueno}, {Tomida}, {Nakahira}, {Kimura}, {Ishikawa}, {Nakagawa}, {Sugizaki},
  {Serino}, {Shidatsu}, {Sugimoto}, {Takagi}, {Matsuoka}, {Arimoto}, {Yoshii},
  {Tachibana}, {Ono}, {Fujiwara}, {Yoshida}, {Sakamoto}, {Kawakubo}, {Ohtsuki},
  {Tsunemi}, {Imatani}, {Tanaka}, {Masumitsu}, {Ueda}, {Kawamuro}, {Hori},
  {Tsuboi}, {Kanetou}, {Yamauchi}, {Itoh}, {Yamaoka}, \&
  {Morii}}]{Nakajima2015}
{Nakajima}, M., {Mihara}, T., {Negoro}, H., {et~al.} 2015, ATEL, 7685

\bibitem[{Negueruela {et~al.}(1999)Negueruela, Roche, Fabregat, \&
  Coe}]{Negueruela1999}
Negueruela, I., Roche, P., Fabregat, J., \& Coe, M.~J. 1999, \mnras, 307, 695

\bibitem[{{Nishimura}(2014)}]{Nishimura2014}
{Nishimura}, O. 2014, \apj, 781, 30

\bibitem[{{Okazaki} {et~al.}(2013){Okazaki}, {Hayasaki}, \&
  {Moritani}}]{Okazaki2013}
{Okazaki}, A.~T., {Hayasaki}, K., \& {Moritani}, Y. 2013, \pasj, 65
  [\eprint[arXiv]{1211.5225}]

\bibitem[{{Pottschmidt} {et~al.}(2005){Pottschmidt}, {Kreykenbohm}, {Wilms},
  {Coburn}, {Rothschild}, {Kretschmar}, {McBride}, {Suchy}, \&
  {Staubert}}]{Pottschmidt2005}
{Pottschmidt}, K., {Kreykenbohm}, I., {Wilms}, J., {et~al.} 2005, \apjl, 634,
  L97

\bibitem[{Poutanen {et~al.}(2013)Poutanen, Mushtukov, Suleimanov, Tsygankov,
  Nagirner, Doroshenko, \& Lutovinov}]{Poutanen2013}
Poutanen, J., Mushtukov, A.~A., Suleimanov, V.~F., {et~al.} 2013, \apj, 777,
  115

\bibitem[{{Raguzova} \& {Lipunov}(1998)}]{Raguzova1998}
{Raguzova}, N.~V. \& {Lipunov}, V.~M. 1998, \aap, 340, 85

\bibitem[{{Roques} {et~al.}(2003){Roques}, {Schanne}, {von Kienlin},
  {Kn{\"o}dlseder}, {Briet}, {Bouchet}, {Paul}, {Boggs}, {Caraveo},
  {Cass{\'e}}, {Cordier}, {Diehl}, {Durouchoux}, {Jean}, {Leleux}, {Lichti},
  {Mandrou}, {Matteson}, {Sanchez}, {Sch{\"o}nfelder}, {Skinner}, {Strong},
  {Teegarden}, {Vedrenne}, {von Ballmoos}, \& {Wunderer}}]{Roques03}
{Roques}, J.~P., {Schanne}, S., {von Kienlin}, A., {et~al.} 2003, \aap, 411,
  L91

\bibitem[{{Sartore} {et~al.}(2015){Sartore}, {Jourdain}, \&
  {Roques}}]{Sartore2015}
{Sartore}, N., {Jourdain}, E., \& {Roques}, J.~P. 2015, \apj, 806, 193

\bibitem[{{Sch{\"o}nherr} {et~al.}(2007){Sch{\"o}nherr}, {Wilms}, {Kretschmar},
  {Kreykenbohm}, {Santangelo}, {Rothschild}, {Coburn}, \&
  {Staubert}}]{Schoenherr2007}
{Sch{\"o}nherr}, G., {Wilms}, J., {Kretschmar}, P., {et~al.} 2007, \aap, 472,
  353

\bibitem[{{Shakura} \& {Sunyaev}(1973)}]{Shakura1973}
{Shakura}, N.~I. \& {Sunyaev}, R.~A. 1973, \aap, 24, 337

\bibitem[{{Skinner} \& {Connell}(2003)}]{Skinner03}
{Skinner}, G. \& {Connell}, P. 2003, \aap, 411, L123

\bibitem[{Staubert {et~al.}(2007)Staubert, Shakura, Postnov, Wilms, Rothschild,
  Coburn, Rodina, \& Klochkov}]{Staubert2007}
Staubert, R., Shakura, N.~I., Postnov, K., {et~al.} 2007, \aap, 465, L25

\bibitem[{{Stella} {et~al.}(1985){Stella}, {White}, {Davelaar}, {Parmar},
  {Blissett}, \& {van der Klis}}]{stella1985}
{Stella}, L., {White}, N.~E., {Davelaar}, J., {et~al.} 1985, \apjl, 288, L45

\bibitem[{{Suchy}(2011)}]{Suchy2011}
{Suchy}, S. 2011, PhD thesis, University of California, San Diego

\bibitem[{{Takeshima} {et~al.}(1994){Takeshima}, {Dotani}, {Mitsuda}, \&
  {Nagase}}]{Takeshima1994}
{Takeshima}, T., {Dotani}, T., {Mitsuda}, K., \& {Nagase}, F. 1994, \apj, 436,
  871

\bibitem[{{Tanaka}(1983)}]{Tanaka1983}
{Tanaka}, Y. 1983, \iaucirc, 3891

\bibitem[{{Terrell} \& {Priedhorsky}(1984)}]{terrell1984}
{Terrell}, J. \& {Priedhorsky}, W.~C. 1984, \apjl, 285, L15

\bibitem[{Tsygankov {et~al.}(2006)Tsygankov, Lutovinov, Churazov, \&
  Sunyaev}]{Tsygankov2006}
Tsygankov, S.~S., Lutovinov, A.~A., Churazov, E.~M., \& Sunyaev, R.~A. 2006,
  \mnras, 371, 19

\bibitem[{Tsygankov {et~al.}(2010)Tsygankov, Lutovinov, \&
  Serber}]{Tsygankov2010}
Tsygankov, S.~S., Lutovinov, A.~A., \& Serber, A.~V. 2010, \mnras, 401, 1628

\bibitem[{Ubertini {et~al.}(2003)Ubertini, Lebrun, di~Cocco, Bazzano, Bird,
  Broenstad, Goldwurm, La~Rosa, Labanti, Laurent, Mirabel, Quadrini, Ramsey,
  Reglero, Sabau, Sacco, Staubert, Vigroux, Weisskopf, \& Zdziarski}]{ibis}
Ubertini, P., Lebrun, F., di~Cocco, G., {et~al.} 2003, \aap, 411, L131

\bibitem[{{Vedrenne} {et~al.}(2003){Vedrenne}, {Roques}, {Sch{\"o}nfelder},
  {Mandrou}, {Lichti}, {von Kienlin}, {Cordier}, {Schanne}, {Kn{\"o}dlseder},
  {Skinner}, {Jean}, {Sanchez}, {Caraveo}, {Teegarden}, {von Ballmoos},
  {Bouchet}, {Paul}, {Matteson}, {Boggs}, {Wunderer}, {Leleux},
  {Weidenspointner}, {Durouchoux}, {Diehl}, {Strong}, {Cass{\'e}}, {Clair}, \&
  {Andr{\'e}}}]{Vedrenne03}
{Vedrenne}, G., {Roques}, J.-P., {Sch{\"o}nfelder}, V., {et~al.} 2003, \aap,
  411, L63

\bibitem[{{Waters} {et~al.}(1989){Waters}, {de Martino}, {Habets}, \&
  {Taylor}}]{Waters1989}
{Waters}, L.~B.~F.~M., {de Martino}, D., {Habets}, G.~M.~H.~J., \& {Taylor},
  A.~R. 1989, \aap, 223, 207

\bibitem[{{Whitlock}(1989)}]{Whitlock1989}
{Whitlock}, L. 1989, \apj, 344, 371

\bibitem[{{Wilson} {et~al.}(2008){Wilson}, {Finger}, \&
  {Camero-Arranz}}]{wilson2008}
{Wilson}, C.~A., {Finger}, M.~H., \& {Camero-Arranz}, A. 2008, \apj, 678, 1263

\bibitem[{{Zhang} {et~al.}(2005){Zhang}, {Qu}, {Song}, \& {Torres}}]{zhang2005}
{Zhang}, S., {Qu}, J., {Song}, L., \& {Torres}, D.~F. 2005, \apjl, 630, L65

\end{thebibliography}

\end{document}